\documentclass[conference]{IEEEtran}
\IEEEoverridecommandlockouts
\usepackage{cite}
\usepackage{amsmath,amssymb,amsfonts}
\usepackage{algorithmic}
\usepackage{graphicx}
\usepackage{textcomp}
\usepackage{xcolor}
\usepackage{graphicx}
\usepackage{subcaption}
\def\BibTeX{{\rm B\kern-.05em{\sc i\kern-.025em b}\kern-.08em
    T\kern-.1667em\lower.7ex\hbox{E}\kern-.125emX}}
\begin{document}

\title{Container Data Item: An Abstract Datatype for Efficient Container-based Edge Computing\\
\thanks{This research is supported by a grant from National Science Foundation.}
}

\author{
\IEEEauthorblockN{1\textsuperscript{st} Md Rezwanur Rahman}
\IEEEauthorblockA{\textit{Computer Science Dept.} \\
\textit{University of Colorado Boulder}\\
Boulder, USA \\
mdra7255@colorado.edu}
\and
\IEEEauthorblockN{2\textsuperscript{nd} Tarun Annapareddy}
\IEEEauthorblockA{\textit{Computer Science Dept.} \\
\textit{University of Colorado Boulder}\\
Boulder, USA \\
tarun.annapareddy@colorado.edu}
\and
\IEEEauthorblockN{3\textsuperscript{rd} Shirin Ebadi}
\IEEEauthorblockA{\textit{Computer Science Dept.} \\
\textit{University of Colorado Boulder}\\
Boulder, USA \\
shirin.ebadi@colorado.edu}
\and
\IEEEauthorblockN{4\textsuperscript{th} Varsha Natarajan }
\IEEEauthorblockA{\textit{Computer Science Dept.} \\
\textit{University of Colorado Boulder}\\
Boulder, USA \\
varsha.natarajan@colorado.edu}
\and
\IEEEauthorblockN{5\textsuperscript{th} Adarsh Srinivasan}
\IEEEauthorblockA{\textit{Computer Science Dept.} \\
\textit{University of Colorado Boulder}\\
Boulder, USA \\
adarsh.srinivasan@colorado.edu}
\and
\IEEEauthorblockN{6\textsuperscript{th} Eric Keller}
\IEEEauthorblockA{\textit{Computer Science Dept.} \\
\textit{University of Colorado Boulder}\\
Boulder, USA \\
eric.keller@colorado.edu}
\and
\IEEEauthorblockN{7\textsuperscript{th} Shivakant Mishra}
\IEEEauthorblockA{\textit{Computer Science Dept.} \\
\textit{University of Colorado Boulder}\\
Boulder, USA \\
mishras@colorado.edu}
}

\maketitle

\begin{abstract}
We present Container Data Item (CDI), an abstract datatype that allows multiple containers to efficiently operate on a common data item while preserving their strong security and isolation semantics. Application developers can use CDIs to enable multiple containers to operate on the same data, synchronize execution among themselves, and control the ownership of the shared data item during runtime. These containers may reside on the same server or different servers. CDI is designed to support microservice based applications comprised of a set of interconnected microservices, each implemented by a separate dedicated container. CDI preserves the important isolation semantics of containers by ensuring that exactly one container owns a CDI object at any instant and the ownership of a CDI object may be transferred from one container to another only by the current CDI object owner. We present three different implementations of CDI that allow different containers residing on the same server as well containers residing on different servers to use CDI for efficiently operating on a common data item. The paper provides an extensive performance evaluation of CDI along with two representative applications, an augmented reality application and a decentralized workflow orchestrator.
\end{abstract}


\section{Introduction}

A microservice-based architecture structures an application as a collection of loosely coupled (micro)services. Each microservice implements a simple functionality with well-defined APIs. The key attractive feature of a microservice is that it can be developed, deployed, and scaled independently. Containers \cite{noauthor_what_nodate-2} provide a very convenient way to implement microservices \cite{noauthor_what_nodate-1},\cite{liu_microservices_2020} at an edge server or the cloud, and have become the de facto standard for implementing microservices. However, since containers isolate their functionality in their own compute environment, any interactions between microservices entail inter-container communications across different IP spaces, typically by using transport layer protocols such as UDP, TCP, HTTP/HTTPS, messaging queues, or remote procedure calls like gRPC. This approach has two shortcomings. First, communication using transport layer protocols requires developers to worry about relatively low-level networking details and any changes in the use of an underlying protocol, e.g. from TCP to gRPC require them to make significant changes in the application code. Second, communication over a (virtual) network involves data copying and user-kernel context switches resulting in significant performance overhead. Researchers have recently started addressing this second shortcoming by developing shared-memory-based inter-container communication mechanisms \cite{bourhim_inter-container_2019, su_pipedevice_2022}. However, these approaches may open up a security loophole, wherein two or more containers having access to a shared memory can potentially interfere with one another, thus violating the strict isolation guarantee. 


In this paper, we propose the Container Data Item (CDI), an abstract datatype at the language runtime level that allows different containers to operate on the same data efficiently and transparently and preserves the important container isolation semantics. A common computing style in a microservice-based IoT application is that an input data object passes through a sequence of microservices, getting operated on by each microservice in turn. For example, in an augmented reality application discussed in detail in Section~\ref{sec:example}, frames in a video stream pass through frame extractor, object detector, and frame combinator containers. Current implementations incur significant data movement cost \cite{farkas_impact_nodate, bziuk_http_2018}, wherein each microservice in the sequence copies a data item from a database or a message queue into local memory, operates on the data item in local memory, and writes the updated data item from local memory to a database or a message queue. Furthermore, any changes in the mechanism used to implement data movement require changes in the application code.

CDI is specifically designed to support such applications. It provides transparency and efficiency by abstracting away the implementation details of data management (data sharing, data movement, etc) and utilizing efficient suitable underlying communication mechanisms that minimize data movement and user-kernel context switches. For example, if the interacting containers are running on the same server, CDI ensures zero-copy \cite{noauthor_zero-copy_2024} and no user-kernel context switching. It provides transparency by ensuring that the application-level code remains the same irrespective of whether the interacting containers are on the same server or different servers. Low-level implementation details are managed transparently by the runtime system. Finally, each container in the sequence is still completely isolated from other application containers.

To demonstrate the utility of CDI, we provide three different implementations, CDI-SHM, CDI-RPC, and CDI-RDMA.
CDI-SHM uses Unix System V \cite{noauthor_unix_2024} shared memory and is applicable when interacting containers are on the same server. CDI-RPC and CDI-RDMA are applicable when interacting containers are on different servers. CDI-RPC uses gRPC as the underlying communication protocol while CDI-RDMA uses RDMA \cite{noauthor_rdma_nodate} (Remote Direct Memory Access) for memory copy across different servers.

Since container-based microservice applications often run within a container orchestration framework such as Kubernetes \cite{noauthor_production-grade_nodate}, we have integrated CDI with Kubernetes. We specify precise CDI semantics ensuring the isolation guarantee of the containers and describe their detailed design and implementation. We have performed an extensive performance evaluation, comparing CDI-based implementations with TCP and gRPC-based implementations. Our evaluation shows significant performance improvement in communication latency over TCP and gRPC, and light reduction in CPU and memory usage. Further, we demonstrate the utility of CDI by building two applications, an augmented reality application where we compare the performance of a CDI-based implementation with a TCP-based and a gRPC-based implementation, and a decentralized workflow orchestrator, where we compare the performance of a CDI-based implementation with Orkes~\cite{noauthor_modern_nodate}, an enterprise-grade conductor platform. The main contributions of this paper are as follows:

\begin{itemize}
    \item CDI moves the support for container interactions from current I/O based methods to the language runtime level, providing transparency and encapsulation, and improving software modularity, reusability, maintainability, and interoperability. To the best of our knowledge, this is the first such effort.
    \item Our prototype implementation demonstrates a significant reduction in latency and resource usage and an increase in throughput.
    \item We demonstrate the utility of CDI by building two popular applications, an augmented reality application, and a decentralized workflow orchestrator.
\end{itemize}

\section{Related Work}

At present, interaction between containers in a microservice based architecture is primarily implemented using network protocols that work across IP spaces (TCP, UDP, HTTP, RPC, etc.) or message queues implemented using these protocols (RabbitMQ~\cite{noauthor_rabbitmq_nodate}, Apache Kafka~\cite{noauthor_apache_nodate}, Apache Pulsar~\cite{noauthor_apache_nodate-1}, etc.). Due to the inherent overhead in these protocols (data copying, context switching, etc), exploring techniques for efficient inter-container communication is an active area of research in both academia and industry. Shimmy \cite{abranches_shimmy_nodate, khasgiwale_shimmy_2023} presents a high-performance inter-container communication mechanism via shared memory channels targeting cloud and edge computing scenarios. Shimmy leverages shared memory for local communication and RDMA synchronization for remote interactions.
It provides significantly higher throughput and lower latency than TCP, UDP or message queues. 

Ubaid et al.~\cite{abbasi_performance_2019} have analyzed the inter-container network bandwidth and compute utilization on an RDMA-enabled Kubernetes cluster, comparing various networking frameworks and observed significant advantages over traditional networks for live migration. Xue et al.~\cite{xue_fast_2019} have shown how deep neural networks can greatly benefit from using an RDMA based system rather than the traditional systems which rely on gRPC on TCP. Fent et al.~\cite{fent_low-latency_2020} have extensively compared TCP with a RDMA or a shared memory based solution for a database management system. This work shows that RDMA based solutions provide an order of magnitude better performance than traditional TCP based approaches. The paper also shows that this performance improvement comes due to lesser context switches in the kernel.

Kun et al. ~\cite{suo_analysis_2018} has published a comprehensive comparison of various container networking technologies, analyzing their performance degradation and overheads in a cloud environment. 
Qualitative comparisons of levels of security, isolation, and overhead and quantitative comparisons of throughput, latency, scalability, and startup cost were made involving various container networks in a realistic cloud environment. They found that a virtualized network in containers incurs non-negligible overhead compared to physical networks. Gerald et al.~\cite{budigiri_network_2021} have benchmarked various network policies and CNI plugins in Kubernetes and analyzed their applicability for 5G low-latency applications. Memif ~\cite{noauthor_fdio_nodate} is another open-source implementation of an efficient inter-container shared memory networking library using the Vector Packet Processing (VPP) technology.

Dhmem \cite{hobson_shared-memory_2021} is a software library designed to manage shared memory across containerized workflow tasks with minimal code changes and performance overhead. It facilitates efficient cross-container communication by leveraging shared memory for local interactions and supports scalable performance and runtime communication configuration. Its design principles aim to simplify integration into existing workflows, providing a unified model for data sharing and synchronization that addresses the challenges posed by containerization in scientific computing environments. A comparison with traditional message serialization methods and a thread-based approach highlights its superior performance in terms of throughput and latency across various workflow configurations, including simple, pipeline, and scatter-gather types.

FreeFlow \cite{kim_freeflow_nodate} is a software-based virtual RDMA networking framework designed to bridge the gap between containerization and
RDMA technologies in cloud-based applications. Its primary goal is to enable containerized applications running in shared cloud
environments to harness the high-performance benefits of RDMA, while simultaneously ensuring the critical properties of isolation, portability, and controllability that are essential for efficient container management. FreeFlow does not require specialized hardware or hardware-based I/O virtualization. Instead, it achieves this by employing a software virtual switch on each server, offering transparent integration with applications running inside containers and resulting in throughput and latency comparable to bare-metal RDMA. 

Finally, our work on CDI is motivated by Resilient Distributed Datasets (RDD) \cite{zaharia_resilient_nodate} which is an abstraction at the core of Spark \cite{apache_spark}. It is a read-only multi-set of data items distributed over a cluster of machines maintained in a fault-tolerant way. An RDD is immutable as the original data does not get changed when multiple workers are working on the data, they rather get a copy of the data to transform. CDI is a similar abstraction where the goal is to minimize data movement when multiple containers operate on them and abstract away the complexities of the underlying networking protocols.

\section{Container Data Item}


CDI is designed to allow multiple containers to operate on a common data item. At a high level, a CDI object can be seen as consisting of three attributes, {\it key}, an id that uniquely identifies the object, {\it data}, that stores the data item, and {\it dataSize}, the size of the data item. In particular, the CDI runtime system provides three abstract data types, {\it CDI} that refers to a CDI object, {\it CDI\_key} to uniquely identify CDI objects, and {\it CDI\_container} to uniquely identify application containers. The unique identities provided by {\it CDI\_key} and {\it CDI\_container} are only within the context of the set of the application containers that comprise an application. For simplicity, the size of a CDI object ({\it dataSize}) at present is fixed when the object is created, and it remains constant throughout the lifetime of the object. See Section~\ref{sec:discussion} for a discussion on variable-sized CDI objects.

A {\it CDI\_key} is simply a string and a {\it CDI\_container} is simply an integer. The actual internal representation of a {\it CDI\_key} is implementation dependent. For example, if the containers sharing the object are on the same server, {\it CDI\_key} would refer to {\it key\_t} if System V shared-memory is used or a filename if POSIX shared-memory is used. Developers may assign a CDI object key either manually, e.g. by reading from a file or using a helper function {\it CDI\_key\_create( )}. The {\it CDI\_container} id is assigned to an application container during application configuration.

The set of operations for a CDI object includes operations to create and destroy a CDI object, manipulate its value, and manage its access control. At any instant, each CDI object is {\it owned} by exactly one container, which can manipulate its value, manage its access control, and destroy the object. No container other than a CDI object's owner has any access to that object, i.e. no other container can read that object's value, update its value, or change its access rights. Further refinement of read/write access rights is discussed in Section~\ref{sec:discussion}.

\begin{figure}
    \centering
\includegraphics[width = \columnwidth]{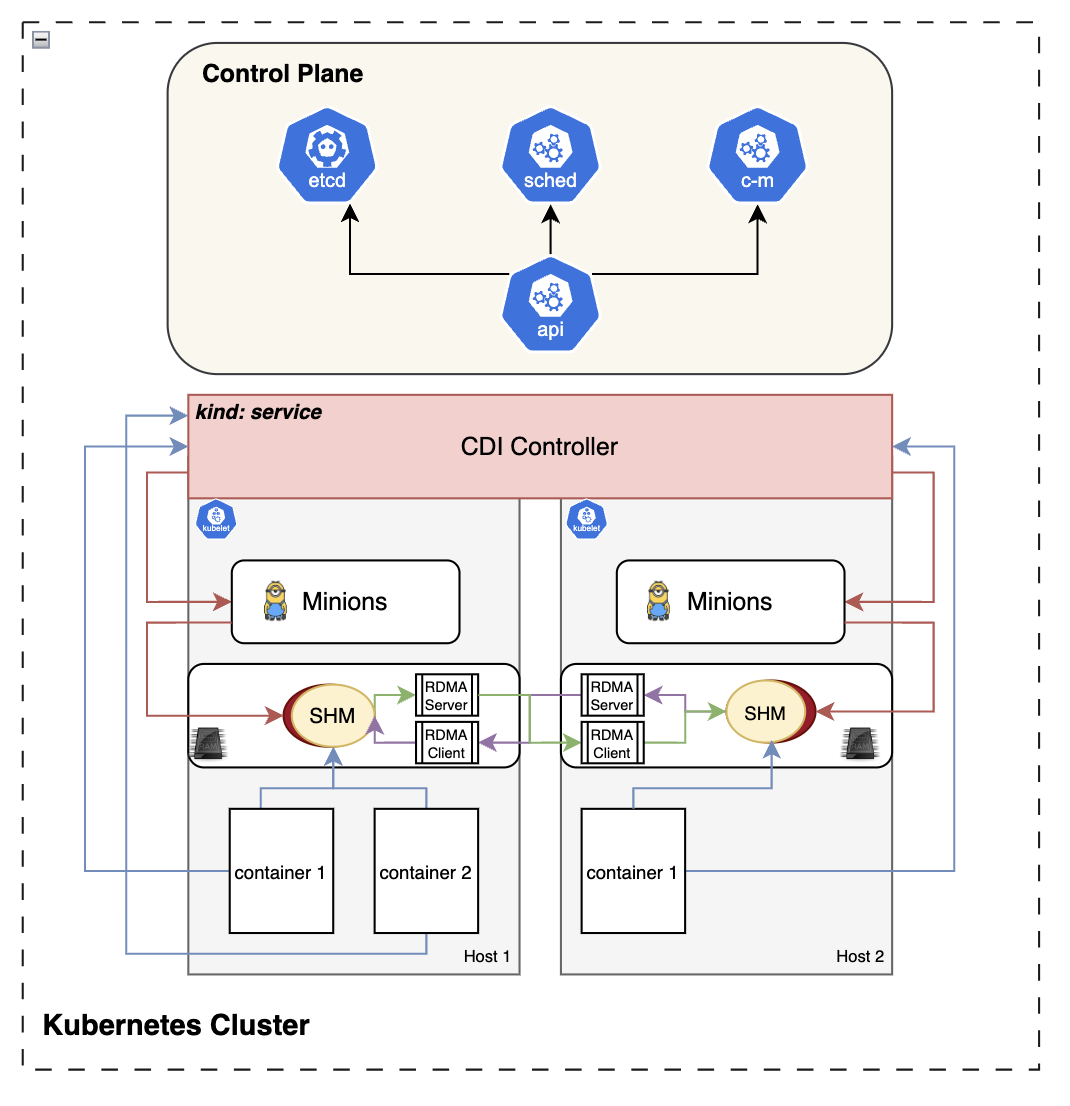}
\caption{CDI Architecture}
    \label{fig:cdi-arch}
\end{figure}

\subsection{CDI Semantics}

\begin{itemize}
    \item A CDI object is created by a container and that container is the owner of that object when created.

    \item A CDI object is owned by exactly one container at any time.

    \item Only the owner of a CDI object can manipulate (read, write, or destroy) the object.

    \item The owner of a CDI object can create a new CDI object that is an exact copy of the original CDI object. Once created, the new CDI object is owned by the container that created that object and is completely independent of any other CDI objects.

    \item The owner C1 of a CDI object can transfer the ownership of that object to another container C2. In one atomic operation, C1 ceases the ownership and C2 acquires the ownership of the object.
\end{itemize}

\subsection{CDI Operations} 

The following operations are supported on a CDI object:

\begin{itemize}
    \item {\it CDI\_create({CDI\_key} k, {\it size\_t} s)}:\\
    - Returns 0 if a CDI object with key {\it k} already exists.\\  
    - Returns -1 if a new CDI object with key {\it k} cannot be created for any other reason such as insufficient memory space.\\
    - Otherwise, creates a new CDI object of data size {\it s} identifiable by key {\it k} and returns 1.\\

    \item {\it CDI\_use(CDI\_key k)}: \\
    - Associates the CDI variable with a CDI object with key {\it k} if that object exists and returns 1.\\
    - Returns 0 if a CDI object with key {\it k} doesn't exist.\\
    
    \item {\it CDI\_copy(CDI c, CDI\_key k )}:\\ 
    - Returns 0 if a CDI object with key {\it k} already exists.\\
    - Returns -1 if a new CDI object cannot be created for any other reason such as insufficient memory space.\\
    - creates a new CDI object identifiable by key {\it k} and returns 1; the new CDI object is an exact copy of the CDI object c.\\
        
    \item {\it CDI\_access( )}: \\
    - Blocks the calling thread until the container has obtained ownership of the CDI object.\\
    
    \item {\it CDI\_transfer({CDI\_container c\_id)})}:\\
    - Transfers the ownership of the CDI object to a different container identifiable by id {\it c\_id}.\\
    
    \item {\it CDI\_destroy( )}:\\
    - Destroys a CDI object.

\end{itemize}

\subsection{CDI Application Example}
\label{sec:example}

\begin{figure*}
    \centering
    \includegraphics[width=0.75\textwidth]{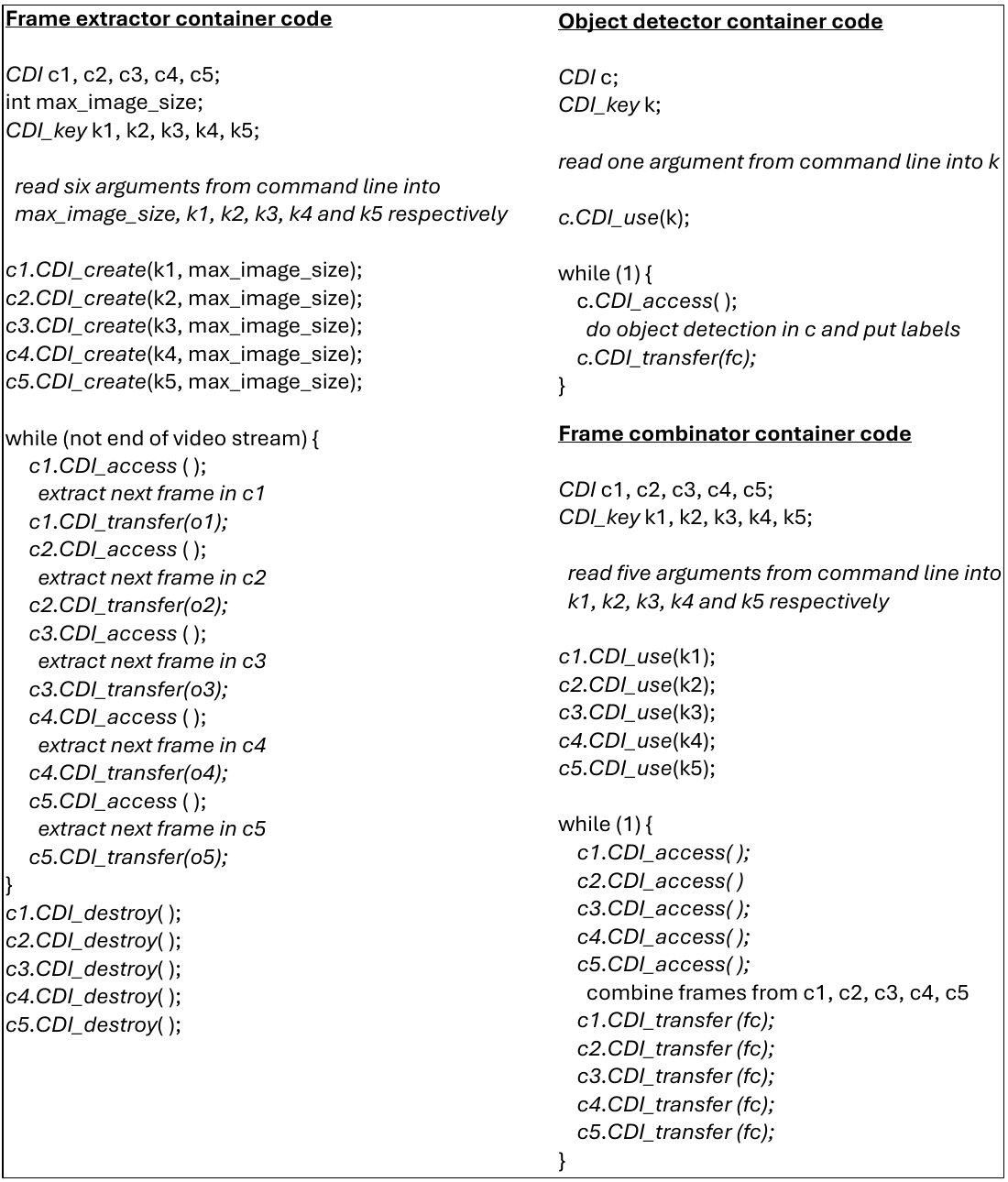}
    \caption{Augmented reality application using CDI (See Section~\ref{sec:example} for explanation).}
    \label{fig:ar_code}
\end{figure*}

To better understand how CDIs are used to build microservice-based applications, we provide a high-level description of an augmented reality application (See Figure~\ref{fig:ar_code}). Later in Section V, we provide a performance evaluation of this application. This application is comprised of three containers performing different roles: {\it frame extractor container}, {\it object detector container} and {\it frame combinator container}. The {\it frame extractor container} creates five CDI objects, c1, ..., c5. For simplicity, we have assumed that the keys for these objects are provided via the command line and are unique. 

In earlier work on building an augmented reality application, we found that the object detector container is the performance bottleneck in this application and we can improve the overall application performance by running multiple instances of this container in parallel, each operating on a different frame~\cite{hu_distributed_2023}. While the optimal number of object detector container instances to run depends on the server capabilities and current load (See~\cite{hu_distributed_2023}), for simplicity, we have chosen to run five instances in this example code. After storing frames in CDI objects c1, ..., c5, the {\it frame extractor container} transfers the ownership of each of these five objects to the five object detector containers respectively.

Each {\it object detector container} waits to get the ownership of its respective CDI object to perform its operations. Once it receives its respective CDI object, it performs object detection and labeling on the frame stored in that object. For our case, we used YOLO-based \cite{redmon_you_2016} real-time object detection on videos. Next, it transfers the ownership of its CDI object to the {\it frame combinator container}, which waits to gain the ownership of the five CDI objects (from each of the five {\it object detector containers}) and then combines these frames to output a video sub-stream. Next, it transfers the ownership of the five CDI objects back to the {frame extractor container}, which reuses these objects to process the next five frames in the video stream.

\section{Design and Implementation}

\subsection{Architecture}
Figure~\ref{fig:cdi-arch} shows the overall architecture of our proposed system. Large-scale container based applications are managed by a container orchestration framework such as Kubernetes~\cite{noauthor_production-grade_nodate} that orchestrates a set of worker nodes within a cluster in the cloud or edge. Each worker node can run one or more Pods, with each Pod hosting one or more containers. Overall, a Kubernetes cluster is comprised of a control plane (API Server, Etcd \cite{noauthor_etcd-ioetcd_2024}, Scheduler and Controller Manager) and the server nodes (Host 1, Host 2, Host 3, ... ). 
The container orchestrator manages the entire infrastructure by deciding what containers to run and which hosts to run them on.
This functionality is exposed via an API to the administrators.



CDI extends a container orchestration framework by introducing two new components, {\it CDI Controller} that provides an interface to the application containers to use CDI objects and {\it CDI minion} that implements CDI objects.
We envision the CDI Controller to be (logically) a part of the orchestrator's control plane. It exposes a set of APIs for the application containers to register themselves, create and destroy CDI objects, and manipulate the ownership of various CDI objects. The CDI Controller is implemented as a Kubernetes service and is accessible to all containers in all nodes. A CDI minion running on a host manages the implementation details of all CDI objects that are being used by some container running on that host.


\subsection{CDI Controller}

The CDI controller exposes an API function for each of the CDI operations, namely {\it CDI\_create()}, {\it CDI\_use()}, {\it CDI\_copy()}, {\it CDI\_access()}, {\it CDI\_transfer()} and {\it CDI\_destroy()}. These functions are invoked by the application containers in response to the corresponding CDI operations invoked by the application. In addition, the CDI controller exposes a {\it CDI\_register(CDI\_container c\_id)} function. During configuration, all application containers that use CDI register with the CDI controller using this API. Unique container id values are provided by the developer for each container and the controller maintains a mapping table of container id, container IP address and server IP address.

\subsection{CDI Minions}

An instance of CDI Minion runs on every node and is
responsible for managing CDI objects’ storage and access
control based on the directives given by the CDI Controller. CDI minions are implemented as a Kubernetes DaemonSet,
which ensures that all current nodes run an instance of a CDI
minion in a pod, a minion pod is added to new nodes joining
a cluster and the minion pod on a node is garbage collected
when that node leaves a cluster. Finally, deleting a DaemonSet
Clean up the minion pods on all nodes.

When an application container invokes an API function of the CDI controller, the controller interacts with the minion(s) running on the relevant application container host(s) to implement the function. To manage CDI objects, a minion maintains a {\it Container Group} for each CDI object, which is a set of containers that can read and manipulate that CDI object, one at a time. In addition, it maintains the identity of the container (one of the containers in the container group) that is the current owner of the object. 
Container Group for a CDI object is created and its owner identity is initialized when the {\it CDI\_create( )} API function is invoked. Membership of this container group is updated when the {\it CDI\_use( )} API function is invoked.

\subsection{CDI Object Storage}

{\it CDI\_create( )} and {\it CDI\_copy( )} result in creation of a new CDI object. On invocation of these functions, the controller contacts the relevant minion(s) to create the new CDI object. For example, when an application container running on a host {\it H1} invokes {\it CDI\_create( )}, the controller contacts the minion on host {\it H1} to create the CDI object. The minion then creates the CDI object in the host’s shared memory. 

If a CDI object is shared between containers running on the same host, we use System V shared-memory for object storage that is created by the minion running on the host and is shared with the container that is the current owner of the object. 
However, sharing memory between containers breaks the barrier of the isolation that containerization provides. To address this, the inter-process communication namespace is shared between the container and the host (done with –ipc="host" in Docker, or hostIPC=true in Kubernetes). Then, the container just needs to know the shm\_id of the shared memory, which is passed in as a configuration file.

If a CDI object is shared between containers running on different hosts, we leverage RDMA technology for sharing. RDMA enables direct copying from one region of memory from one host
to another without involving the processor. To enable RDMA, each container runs an instance of an RDMA server and an RDMA client. 

To facilitate RDMA over an Ethernet network within containers, we need to employ InfiniBand over Ethernet (IBoE) HCA. Our current implementation leverages the Mellanox RDMA Shared Device Plugin ~\cite{noauthor_mellanoxk8s-rdma-shared-dev-plugin_2024}, which operates as a DaemonSet. This plugin seamlessly equips containers with the necessary HCA (Host Channel Adapter) resources, enabling efficient RDMA communication.

Finally, when the current owner of a CDI object invokes {\it CDI\_destroy( )}, the controller contacts the appropriate minion(s), which de-allocates the object memory as well as all metadata (owner id, container group, etc) associated with that object.

\subsection{CDI Object Access Control}

CDI Minions manage access to a CDI object, ensuring that only the current owner of the object has the read/write permissions for that object. When the {\it CDI\_transfer( )} API function is invoked for a CDI object, the controller instructs the appropriate minion(s) to update the owner id of the object. The minion(s) then updates the access control of the object and the owner id of the object. This is done only if the API function is invoked by the current owner of the object. If the two containers (container owner and the next owner) are on the same host, access control is changed by simply invoking shmctl( ).

If the two containers (current owner and the next owner) are on different hosts, RDMA is used to first (direct memory) copy the object from one minion (current owner host) to another minion (next minion host). The sequence of steps are: (1) the minion on the current owner host revokes access permission from the current owner; (2) RDMA is used to copy the object; (3) the minion on the current owner host changes the owner id of the object to the next owner; (4) the minion on the next owner host changes the owner id of the object to the next owner; (5) the minion on the next owner host grants access permissions to the next owner. This sequence ensures that a CDI object has at most one owner at any instant.


\section{Evaluation}

\subsection{Setup}
For raw measurements and augmented reality application, we use m400 Cloudlab Nodes \cite{noauthor_cloudlab_nodate} with aarch64 
architecture, dual-port Mellanox ConnectX-3 10 GB NIC (PCIe v3.0), 8 ARMv8 64-bit CPU cores and 65GB Memory
space in our Kubernetes cluster.  We dedicate 1 node to the
master and 3 nodes to the worker. Setup for decentralized workflow orchestrator is discussed in Section~\ref{sec:orchestrator} 

\subsection{Raw Measurements}

We first compare the performance of CDI-SHM and CDI-RDMA with TCP and gRPC. We measure latency for different message sizes. In the context of an application, a message size corresponds to the size of a data item being operated on.
Tables~\ref{tab:RTT_single_host} and ~\ref{tab:RTT_multiple_hosts} show mean latency for respectively single-server (when all containers run on the same server) and multiple servers (when containers run on different servers). We can see that both CDI-SHM and CDI-RDMA outperform TCP or gRPC and this performance gain gets increasingly better as message sizes increase.

We attribute this performance improvement to reduced data copying and context switching in CDI-SHM and CDI-RDMA. Further, no code changes are required in the application whether it is running on CDI-SHM or CDI-RDMA, while code changes are required when moving TCP to gRPC or gRPC to TCP based implementations.

\begin{table}[ht]
    \centering
    \caption{Latency (milliseconds) in a single host configuration.}
    \begin{tabular}{|c|cccc|}
    \hline
         &  \textbf{10 KB} &  \textbf{100 KB} & \textbf{1 MB} & \textbf{10 MB} \\
    \hline
         \textbf{gRPC} & 1.25  & 1.49  & 6.2  & 39.12   \\
         \textbf{TCP}  & 0.13  & 0.19  & 0.69  & 16.9  \\
         \textbf{Shared Memory}  & 0.005 & 0.05 & 0.47 & 4.5 \\
    \hline
    \end{tabular}

    \label{tab:RTT_single_host}
\end{table}

\begin{table}[ht]
    \centering
    \caption{Latency (milliseconds) in a multi-host configuration.}
    \begin{tabular}{|c|cccc|}
    \hline
         &  \textbf{10 KB} &  \textbf{100 KB} & \textbf{1 MB} & \textbf{10 MB} \\
    \hline
         \textbf{gRPC} & 1.34  & 1.63  & 6.05  & 73.93 \\
         \textbf{TCP}  & 0.12  & 0.14  & 0.81  & 17.47 \\
         \textbf{RDMA}  & 0.014  & 0.094  & 0.87 & 8.64 \\
    \hline
    \end{tabular}
    \label{tab:RTT_multiple_hosts}
\end{table}

\subsection{Augmented Reality Application}
\label{sec:ar_performance}

We have implemented the augmented reality application described in Section~\ref{sec:example} in three ways. The first one uses TCP for inter-container communication, the second one uses gRPC for inter-container communication, and the third one uses CDI. For CDI, we used CDI-SHM when all containers run on the same server and CDI-RDMA when different containers run on different servers.

\begin{figure*}[ht]
    \centering
    \subfloat[]{
        \includegraphics[width=0.31\linewidth]{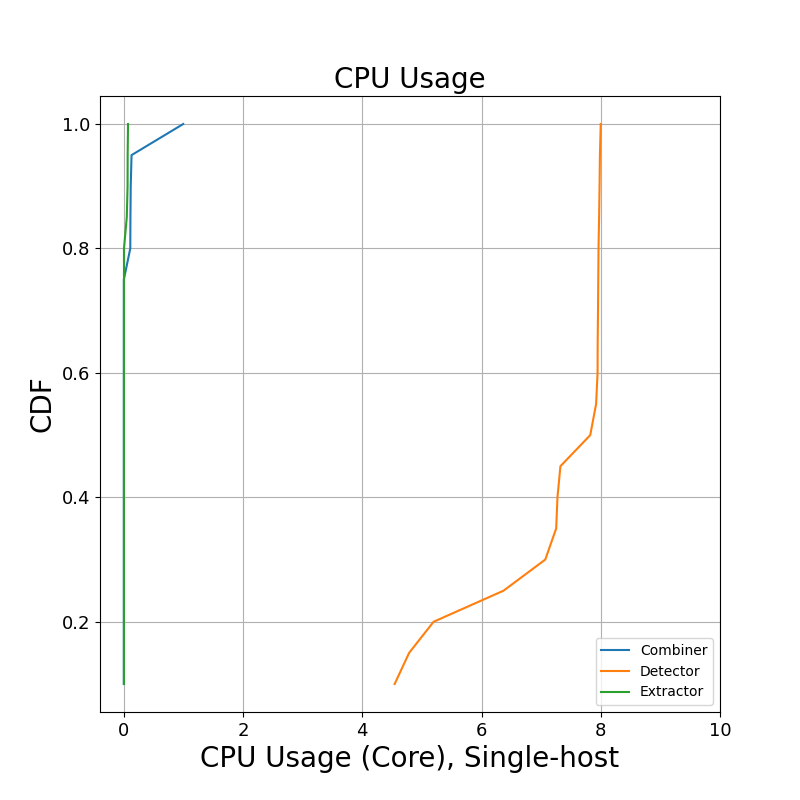}%
        }
    \subfloat[]{
        \includegraphics[width=0.31\linewidth]{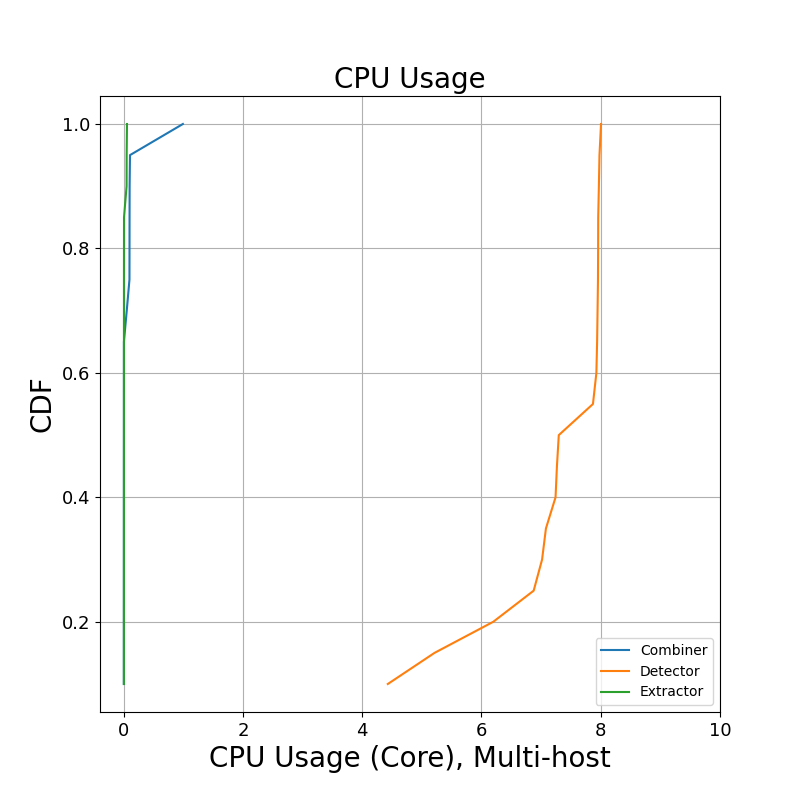}%
        }
    \subfloat[]{%
        \includegraphics[width=0.31\linewidth]{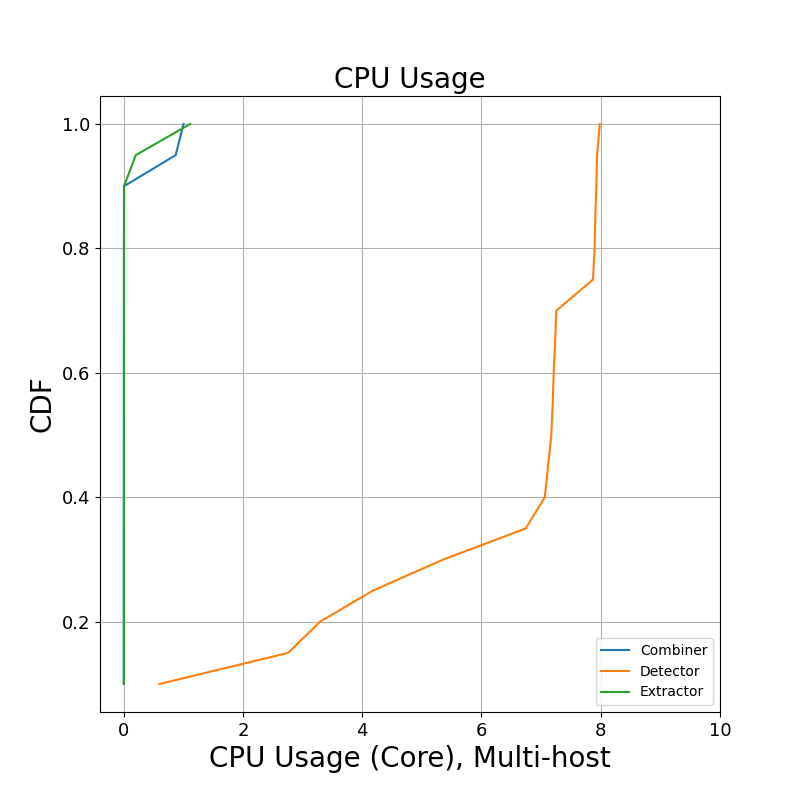}%
        }%
    \\
    \subfloat[]{%
        \includegraphics[width=0.31\linewidth]{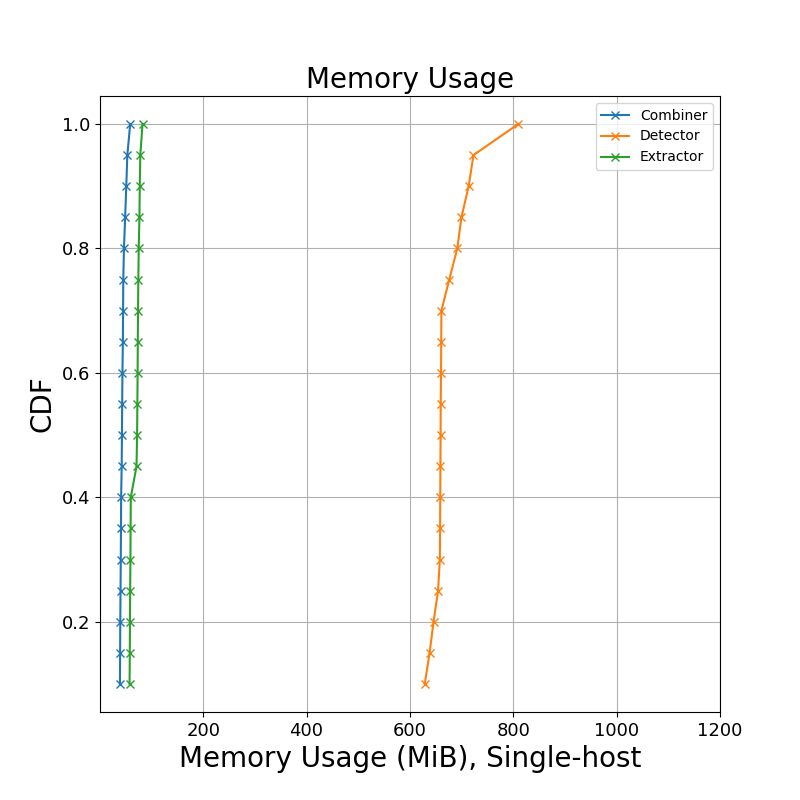}%
        }%
    \subfloat[]{
        \includegraphics[width=0.31\linewidth]{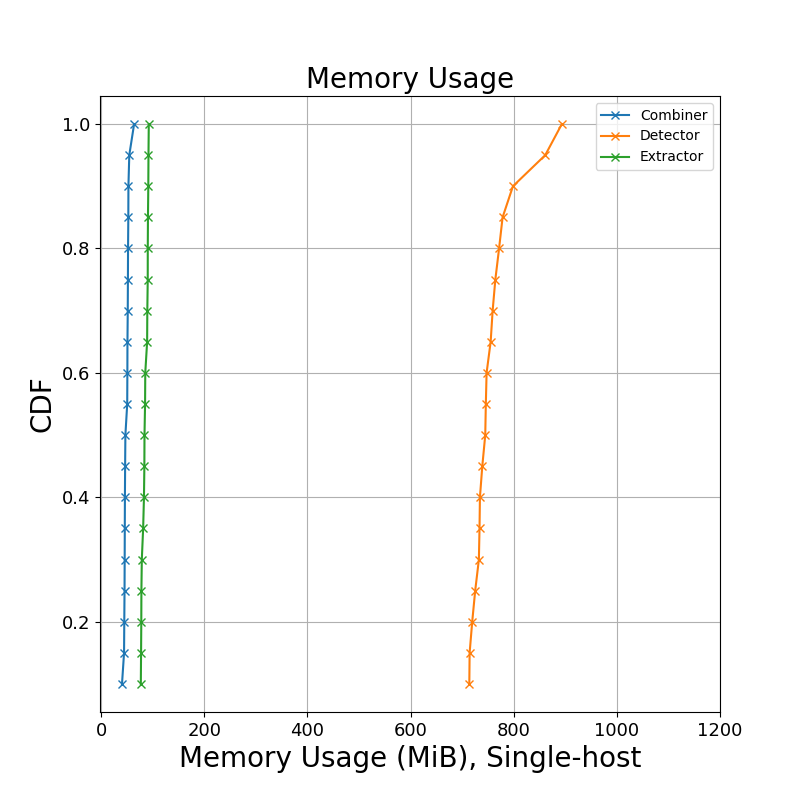}%
        }
    \subfloat[]{
        \includegraphics[width=0.31\linewidth]{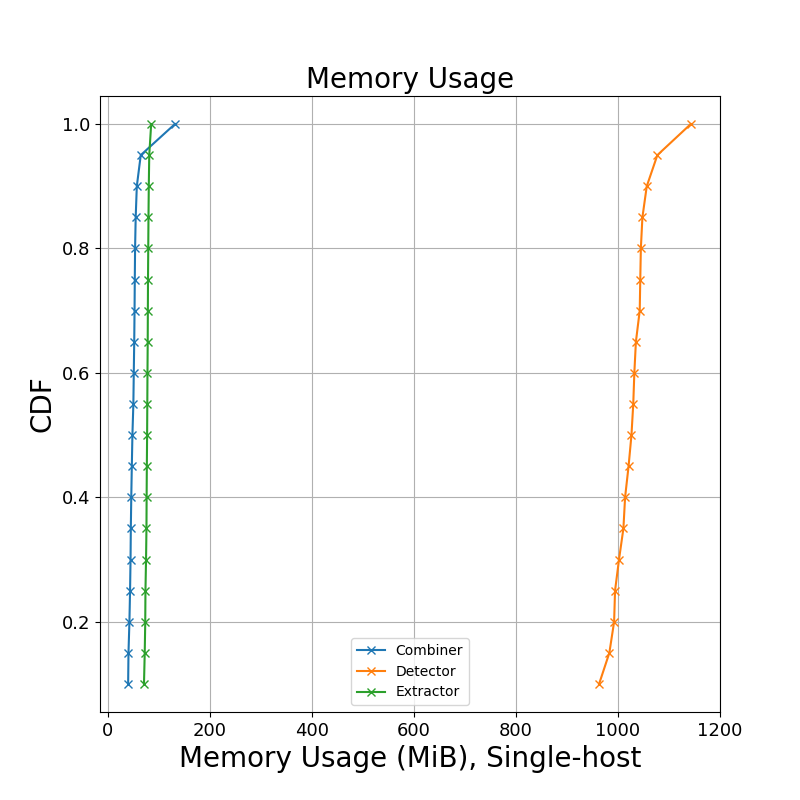}%
        }
    \caption{Single-host setup of Augmented Reality Application. (a), (b) and (c): CPU usage for TCP, gRPC and CDI-SHM based implementations respectively; (d), (e) and (f): Memory usage for TCP, gRPC and CDI-SHM based implementations respectively.}
    \label{fig:ar_singlehost}
\end{figure*}

\begin{figure*}[ht]
    \centering
    \subfloat[]{
        \includegraphics[width=0.31\linewidth]{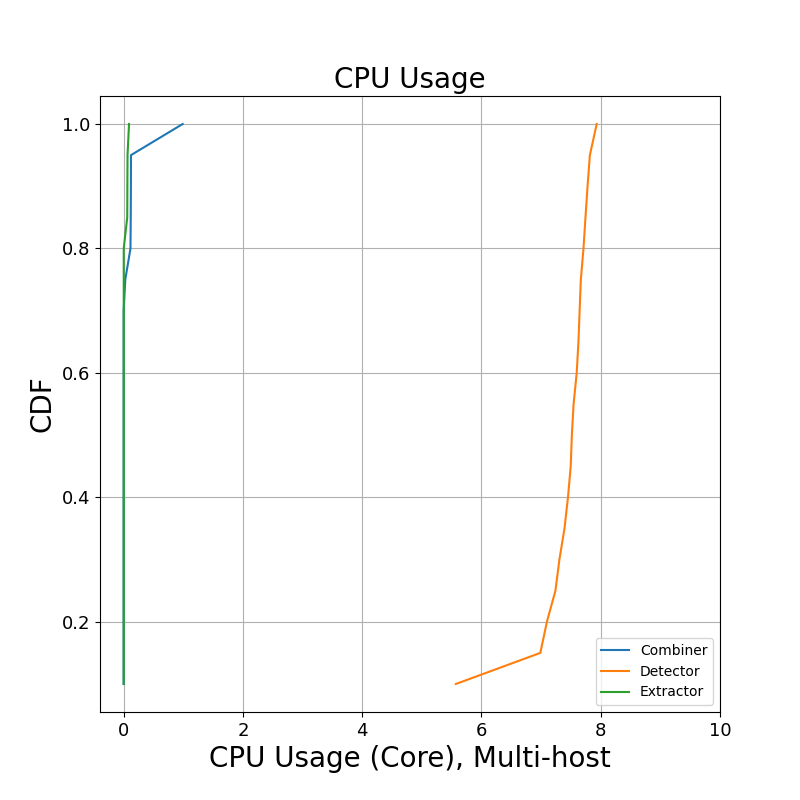}%
        }
    \subfloat[]{
        \includegraphics[width=0.31\linewidth]{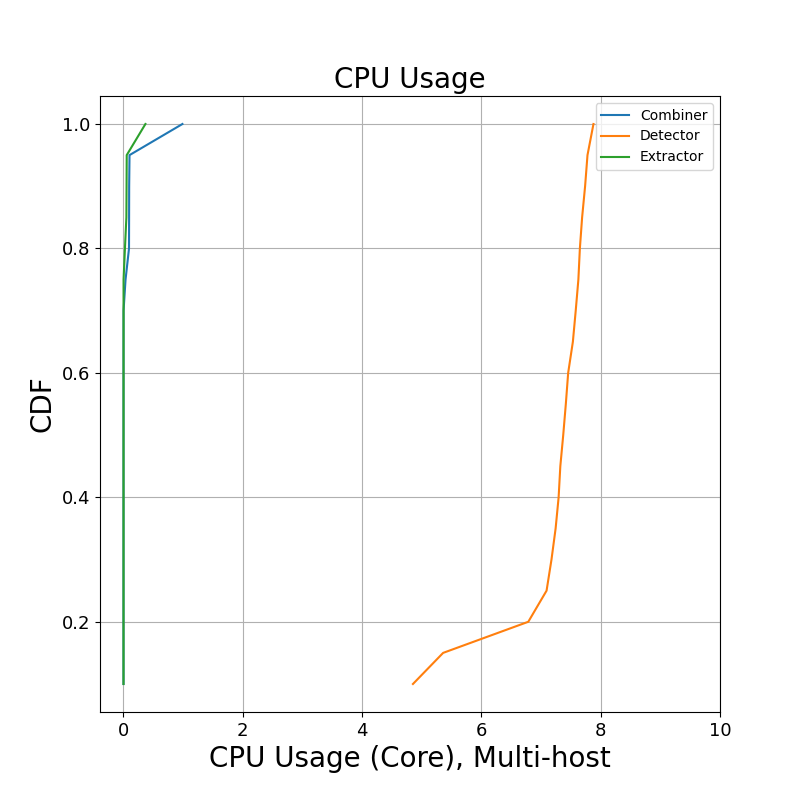}%
        }
            \subfloat[]{%
        \includegraphics[width=0.31\linewidth]{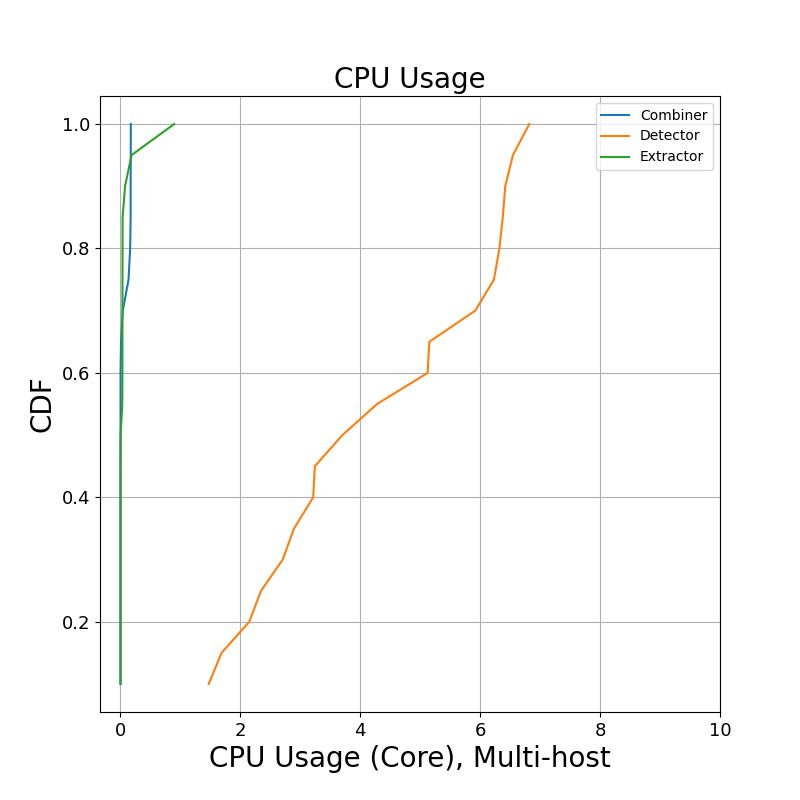}%
        }%

        \subfloat[]{%
        \includegraphics[width=0.31\linewidth]{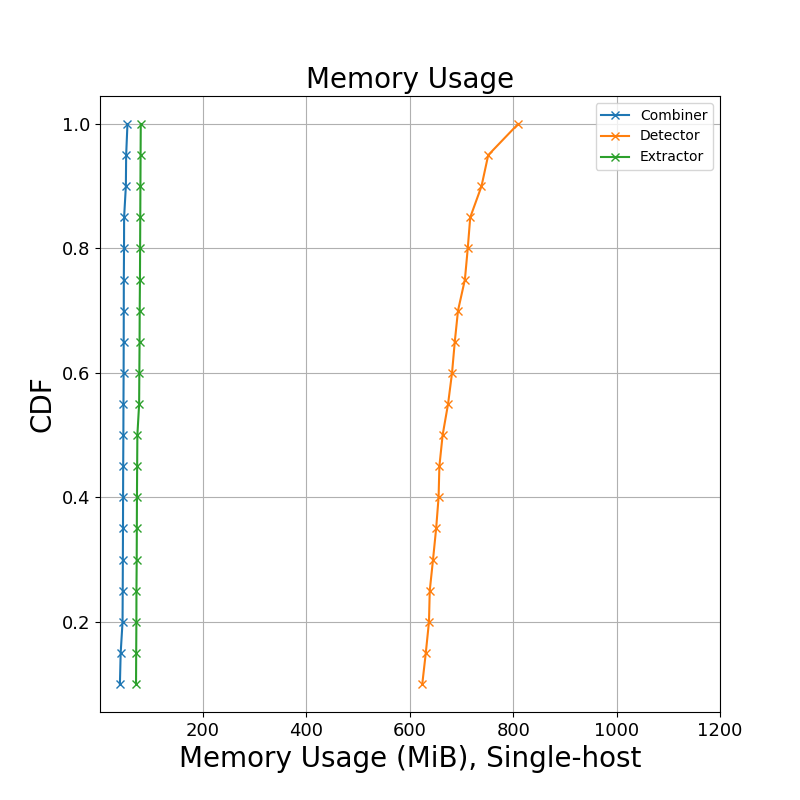}%
        }%
    \subfloat[]{%
        \includegraphics[width=0.31\linewidth]{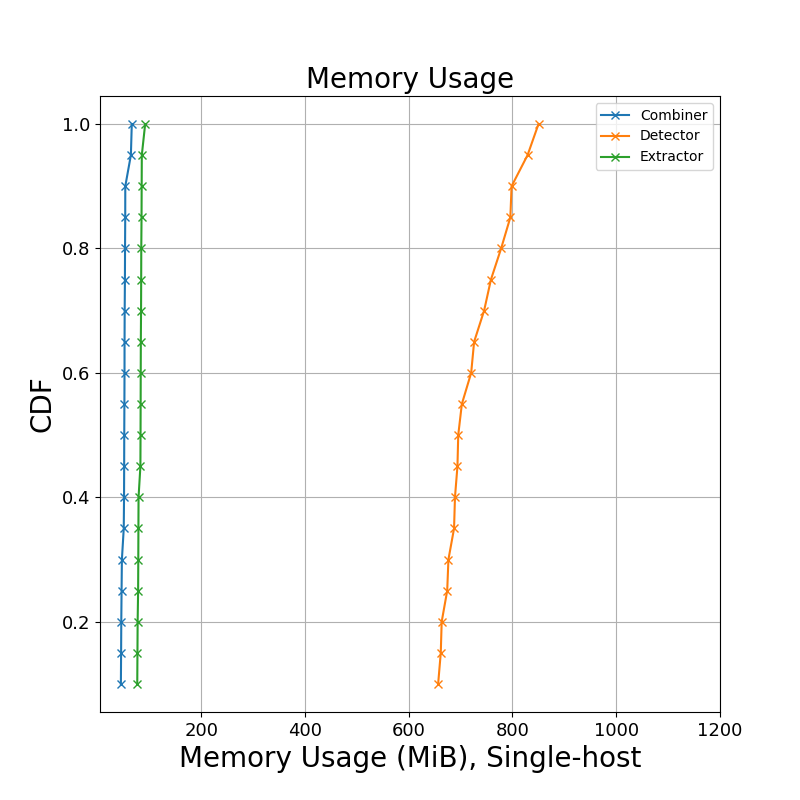}%
        }%
            \subfloat[]{%
        \includegraphics[width=0.31\linewidth]{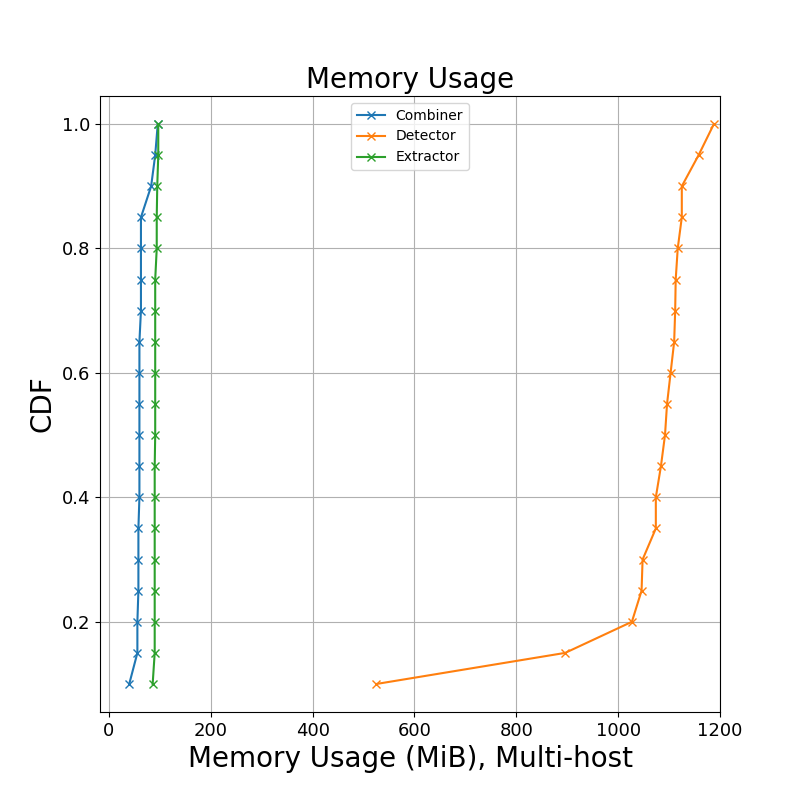}%
        }%
    \caption{Multi-host setup of Augmented Reality Application. (a), (b) and (c): CPU usage for TCP, gRPC and CDI-RDMA based implementations respectively; (d), (e) and (f): Memory usage for TCP, gRPC and CDI-RDMA based implementations respectively. }
    \label{fig:ar_multiplehost}
\end{figure*}

Figures~\ref{fig:ar_singlehost} and \ref{fig:ar_multiplehost} show CPU usage and memory usage in the three implementations for single server and multiple servers respectively. We notice that there isn't any noticeable difference in CPU usage in frame extractor and frame combinator containers among the three implementations (very low CPU usage).
As expected object detector container consumes most of the CPU time and here we see improvement in CPU usage in CDI-based implementations. Again this can be attributed to reduced context switching and data copying in CDI-based implementations. 

Memory usage on the other hand is higher in the CDI-based implementations when compared to TCP or gRPC-based implementations. This points to a source of inefficiency in CDI, which is that the size of a CDI object is fixed when created and it remains fixed throughout the duration of the object. For example, in our experiments, we fixed the CDI object size to 10 MB, while these objects store video frames that were typically around 2 MB. TCP and gRPC-based implementations on the other hand use much less memory for storing these frames. Note that larger memory usage also negatively impacts latency performance in CDI-RDMA or CDI-gRPC implementations. As further discussed in Section~\ref{sec:discussion}, developing a variable-sized CDI is part of our future work.

\subsection{Decentralized Workflow Orchestrator}
\label{sec:orchestrator}

A workflow orchestrator manages the execution of a complex task across multiple nodes in a distributed environment. It is comprised of a {\it workflow controller}, a {\it workflow scheduler}, and several {\it workflow workers}. The workflow controller serves as the interface to the orchestrator, allowing users to register tasks, and define and initiate workflows. The workflow scheduler is responsible for tracking the status of workflows and scheduling tasks across available worker nodes, ensuring efficient execution. The workflow workers are the entities that perform the tasks defined in the workflows. 

For example, consider an image processing application that performs a sequence of operations on a set of images, such as  deblur\_image $\rightarrow$ denoise\_image $\rightarrow$ classify\_image. A workflow orchestrator ensures that the tasks are executed in the correct order on each image optimizing resource utilization. Orchestrators like Netflix Conductor~\cite{blog_netflix_2017}, Temporal~\cite{noauthor_product_nodate}, and  Apache Airflow~\cite{noauthor_home_nodate} execute such pipelines by having workers continuously poll the orchestrator for tasks they are responsible for. The orchestrator maintains a queue for each task and assigns tasks to workers upon request, providing instructions and data typically linked via an object storage service. In this setup, for each task in the sequence, the image is downloaded from the object storage, processed, and then uploaded back. 

The main performance limitation of these current orchestrators is the significant data movement cost, as each task in the sequence involves downloading images from and uploading images to object storage. This results in lower throughput and limits scalability. We address this limitation by using CDI to share intermediate images between tasks in the sequence. More specifically, we have implemented a decentralized workflow orchestrator that uses CDI to manage workflows and incorporates intelligent worker queues for enhanced efficiency. The overall design consists of the following steps:

\begin{enumerate}
    \item Registration: The workflow controller, scheduler, and all workflow workers are containers. They register themselves with the CDI controller using the {\it CDI\_register( )} API. 
    \item Workflow Initialization: When a workflow is initialized, the scheduler downloads the image and creates a CDI object using the {\it CDI\_create( )} API. It then writes the image to the CDI object.
    \item First Task Assignment: The scheduler determines which worker should execute the first task on the image based on available worker metrics such as current CPU load and memory. The ownership of the CDI object is transferred directly to the selected worker using the {\it CDI\_transfer( )} API, while the task instruction is pushed to the respective worker's queue.
    \item Task Execution: Workers continuously poll their queues for tasks. Upon receiving a task, a worker executes it. After completing the task, the worker reports the status to the scheduler.
    
    \item Scheduling Strategy: We have experimented with two scheduling here: 
    \begin{itemize}
        \item {\it CDI (option 1)}: The scheduler always schedules all the tasks of a particular workflow on the same worker, prioritizing affinity. The worker retains ownership of the CDI object, except if the task executed was the last in the workflow, in which case the scheduler regains ownership of the CDI object.
        \item {\it CDI (option 2)}:  The scheduler chooses a worker based on available worker metrics and transfers the ownership of the CDI object to that worker. So, ownership of the CDI object is transferred between workers for every task of the workflow. 
    \end{itemize}
     Once the ownership of data is transferred, the scheduler pushes the new task instruction to that worker's queue.
    
    \item Workflow Completion: When the execution of the workflow is completed, the scheduler regains ownership of the CDI object. It uploads the processed image in the object store and then destroys the CDI object using the {\it CDI\_destroy( )} API.
    
\end{enumerate}

Our decentralized orchestrator uses the Postgres database for managing workflow and task information, and Redis for worker queue. Our approach significantly reduces the data movement costs and optimizes resource utilization, effectively addressing the limitations of traditional orchestrators in handling data-intensive tasks.

\subsubsection{Performance}

For evaluation, we used CDI-RPC to implement our CDI-based decentralized workflow orchestrator and compare its performance with Orkes~\cite{noauthor_modern_nodate}, an enterprise-grade conductor platform. For Orkes, we utilized Orkes-hosted dedicated cluster for the control plane and 2 worker replicas for task execution. A similar setup was followed for our CDI-based workflow orchestrator, where the controller, scheduler, and 2 worker replicas were containerized and deployed. We used CloudLab servers for evaluation, where we created a Kubernetes cluster with 4 m400 servers, each running a different container. 

A workflow in our evaluation is comprised of a single image being processed by the sequence of three tasks mentioned earlier. The image is stored initially in an object storage (AWS S3) and the final processed image is stored back in the object storage. Our evaluation is done for two different scenarios. In both of the scenarios, we used the ``Weather Image Recognition" dataset from Kaggle. This dataset contains 6,862 images of different weather types. The image sizes vary from 10 KB to 4 MB. We uploaded this dataset to the AWS S3 object store and used the links as input to our workflows. In Scenario 1, we ran ten workflows in parallel for same-sized images. We repeated this scenario for image sizes 10 KB, 100 KB, 500 KB, and 2 MB. In Scenario 2,  We experimented with increasing numbers of concurrent workflow executions, 1, 50, 100, 200, and 500 where images were randomly selected from the ``Weather Image Recognition" dataset. In both scenarios, we measured the average latency for processing a single flow, 
and throughput, which is the number of workflows processed per second.

\begin{figure*}
    \centering
\includegraphics[width = \textwidth]{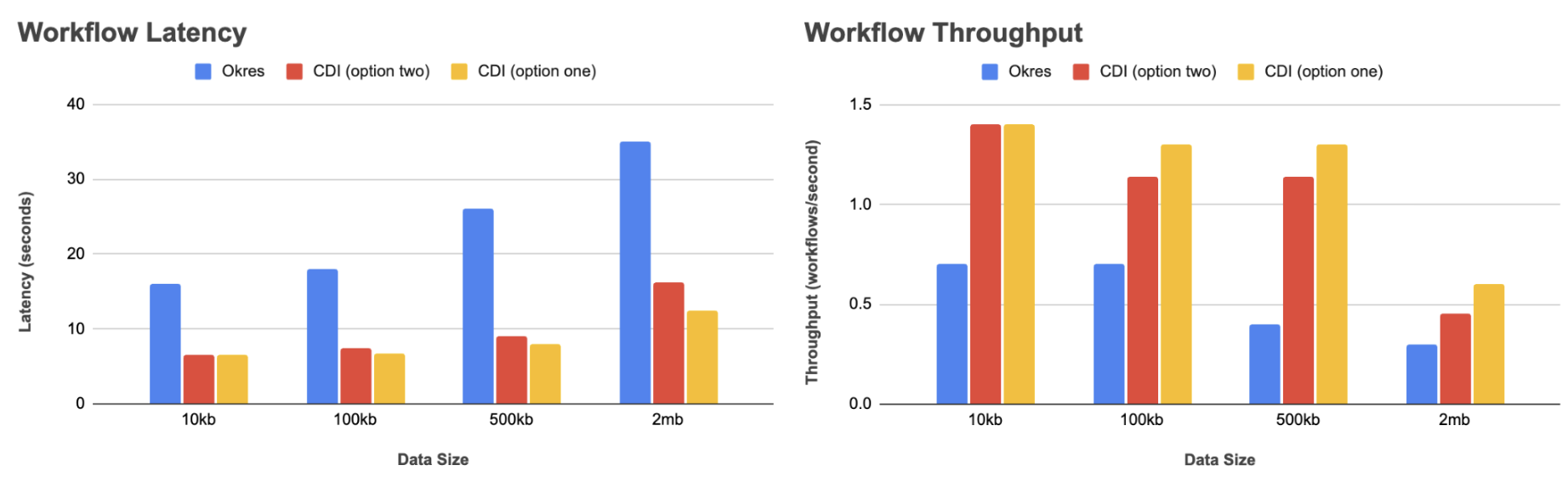}
\caption{Workflow Orchestrator performance for different image sizes; Orkes vs CDI-RPC based implementations.}
    \label{fig:scenario_one}
\end{figure*}

\begin{figure*}
    \centering
\includegraphics[width = \textwidth]{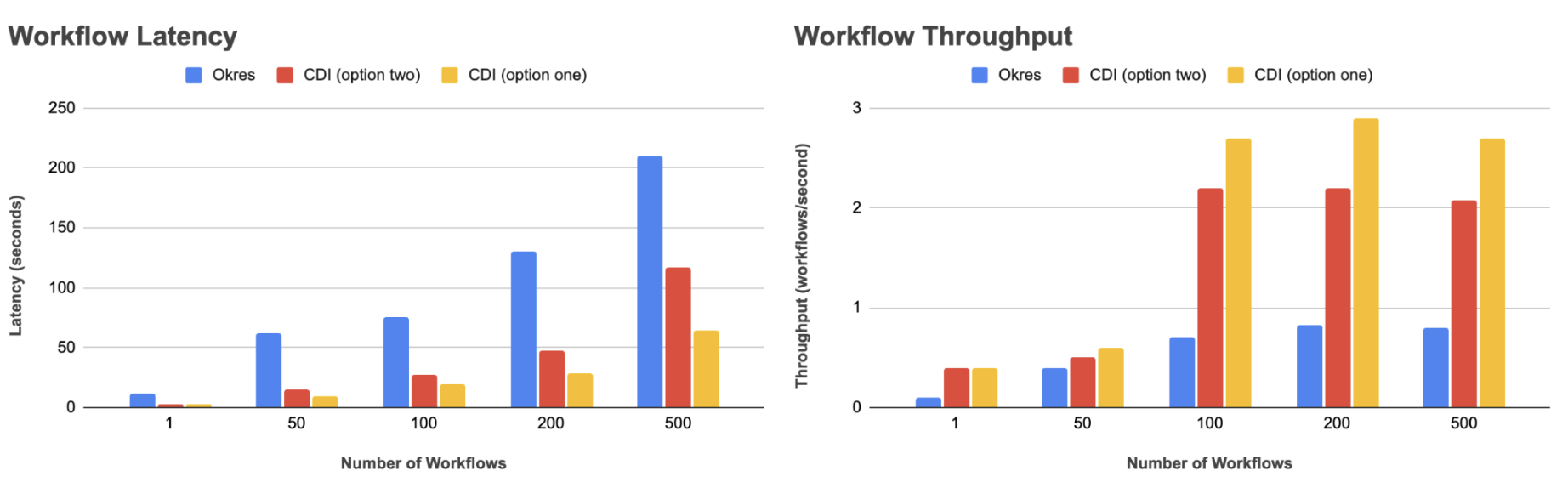}
\caption{Workflow Orchestrator performance for different number of concurrent workflows; Orkes vs CDI-RPC based implementations.}
    \label{fig:scenario_two}
\end{figure*}

Figures~\ref{fig:scenario_one} and \ref{fig:scenario_two} show the performance measured for Scenario 1 and Scenario 2 respectively. We observe that in both scenarios, the CDI-based orchestrator outperforms Orkes for both latency and throughput. 

In Scenario 1, we observe that as the image size increases, the CDI-based orchestrator's performance gets increasingly better compared to Orkes. This is due to low data movement costs in CDI-based orchestrators. In CDI (option 1), where tasks were scheduled on the same worker, performance is slightly better than in CDI (option 2). We believe this is because the ownership of the CDI object doesn't change back and forth between the workers in option 1 and the load on the server is relatively light so the benefit of choosing the best worker based on worker metrics is not much.

Finally, in the second scenario, as the workload increased, the CDI-based orchestrator maintained a more consistent performance compared to Orkes. The efficient scheduling strategy of the CDI-based orchestrator allowed it to handle higher loads more effectively. The throughput remained high, and the latency increase was much less pronounced compared to Orkes.

Overall, our evaluation demonstrates that the CDI-based workflow orchestrator significantly enhances performance by reducing data movement costs and optimizing resource utilization, making it a superior choice for handling data-intensive tasks.

\section{Discussion} \label{sec:discussion}

We have presented Container Data Item (CDI), an abstract datatype that allows multiple containers to operate on the same data item without violating the strong isolation semantics of the containers. We also provide three efficient implementations of CDI, one where all containers are running on the same server and the other two where different containers may run on different servers. A performance comparison with the current practices of using inter-container communications across different IP spaces (TCP, RPC) shows that CDI provides a significant performance improvement in terms of latency and throughput. Finally, we have demonstrated the generality of CDI by constructing two popular IoT applications using CDI resulting in significant performance improvement.

The key contribution of this paper is that CDI elevates data sharing among containers from the current I/O-based methods via TCP, RPC or message queues to the language level where the developers can focus on high-level design issues without getting bogged down by implementation details. As an abstract datatype, CDI encapsulates the data and the operations that can be performed on the data, providing a clear interface and hiding the implementation details, thus leading to more modular and maintainable code. Indeed, developers don't need to know whether CDI has been implemented internally using shared memory or RDMA or I/O-based methods. More importantly, the underlying implementation of CDI can now be changed transparently without any need for application code modification.
    
This paper provides a simple definition of CDI and defines a minimal set of operations for building IoT applications. Our key focus was to ensure that container semantics are preserved. Further investigation is needed to refine this definition and perhaps incorporate additional operations as needed. For example, the size of a CDI object at present is fixed once it is created. We chose this for simplicity. However, a developer may reuse a CDI object for storing different items at different times, e.g. the augmented reality application reuses the same set of CDI objects for different frames over time. While fixed CDI object size simplifies implementation, it requires developers to estimate in advance the maximum size that would be needed for a CDI object when it is created. Further, as we saw in Section~\ref{sec:ar_performance}, this may lead to inefficiency if the item sizes can vary widely and the same CDI object is used to represent them. Defining a variable-sized CDI object, wherein its size may be changed over time is an interesting question for further investigation.

An important guiding principle in defining CDI was to ensure that a CDI object is owned by at most one container at any instant. An interesting question is whether a CDI object access semantics may further be refined. For example, a publish-subscribe type of IoT application would ideally need an object to be read-accessible to multiple containers concurrently. An alternate semantics for a CDI object could be that at any instant, either a single container has (exclusive) read/write access, or multiple containers have concurrent read-only access. The most important issue that needs to be considered here is whether container semantics, particularly its security and privacy are preserved when multiple containers can access (read-only) the same object at the same time. Another issue that needs to be considered is which container among the ones that have read-only access to a CDI object would be the owner that could transfer the access rights of the CDI object. It is critical that concurrent read-only access and access rights transfer do not result in interference between containers.

The paper presents a proof-of-concept prototype that demonstrates the benefits of CDI. At present, it does not include garbage collection to destroy all orphaned CDI objects and reclaim resources held by them. One way, a CDI object may become an orphan is if a developer does not explicitly destroy that object (using {\it CDI\_destroy( )}) after use. Another way would be if no CDI variables refer to a CDI object. Note that the CDI variable reference may be changed using {\it CDI\_use( )}. Incorporating a garbage collector is part of our future work.

CDI may be implemented by any suitable underlying communication technology. In this paper, we have provided implementations using System V shared memory, RDMA and gRPC. In the future, we would like to utilize CXL (Compute Express Link)~\cite{noauthor_compute_2023} for implementing CDI over a cluster of servers connected to a CXL-enabled memory
pool. We expect that this would further reduce latency and increase throughput down to load and store latency bypassing any NIC overhead.

\bibliographystyle{IEEEtran}
\bstctlcite{MyBSTcontrol}
\bibliography{refs}



\end{document}